\documentclass[aps,prfluids,showpacs,notitlepage,superscriptaddress,longbibliography]{revtex4-1}
\usepackage{epsfig,graphics,amssymb,amsmath,subeqnarray,color,bm,bbm}
\usepackage[toc,page]{appendix}
\usepackage{mathtools}
\usepackage{xcolor}
\usepackage{float}
\usepackage{graphicx}
\usepackage[colorlinks=true,linkcolor=blue]{hyperref}

\begin{document}
\title{Efficiency limits of the three-sphere swimmer}
\author{Babak Nasouri}
\affiliation{Max Planck Institute for Dynamics and Self-Organization (MPIDS), 37077 Goettingen, Germany}
\author{Andrej Vilfan}
\email{andrej.vilfan@ds.mpg.de}
\affiliation{Max Planck Institute for Dynamics and Self-Organization (MPIDS), 37077 Goettingen, Germany}
\affiliation{Jo\v{z}ef Stefan Institute, 1000 Ljubljana, Slovenia}
\author{Ramin Golestanian}
\email{ramin.golestanian@ds.mpg.de}
\affiliation{Max Planck Institute for Dynamics and Self-Organization (MPIDS), 37077 Goettingen, Germany}
\affiliation{Rudolf Peierls Centre for Theoretical Physics, University of Oxford, Oxford OX1 3PU, United Kingdom}
\date{\today}

\begin{abstract} 
We consider a swimmer consisting of a collinear assembly of three spheres connected by two slender rods. This swimmer can propel itself forward by varying the lengths of the rods in a way that is not invariant under time reversal. Although any non-reciprocal strokes of the arms can lead to a net displacement, the energetic efficiency of the swimmer is strongly dependent on the details and sequences of these strokes, and also the sizes of the spheres. We define the efficiency of the swimmer using Lighthill's criterion, i.e., the power that is needed to pull the swimmer by an external force at a certain speed, divided by the power needed for active swimming with the same average speed. Here, we determine numerically the optimal stroke sequences and the optimal size ratio of the spheres, while limiting the maximum extension of the rods. Our calculation takes into account both far-field and near-field hydrodynamic interactions. We show that, surprisingly, the three-sphere swimmer with unequal spheres can be more efficient than the equally-sized case. We also show that the variations of efficiency with size ratio is not monotonic and there exists a specific size ratio at which the swimmer has the highest efficiency. We find that the swimming efficiency initially rises by increasing the maximum allowable extension of the rods, and then converges to a maximum value. We calculate this upper limit analytically and report the highest value of efficiency that the three-sphere swimmer can reach.
\end{abstract}
\maketitle
\section{Introduction}
Optimal swimming at small scales has been an immense topic of interest due to both its application in biomedial engineering \cite{nelson2010}, and the possible insight it may provide into evolutionary processes of microorganisms \cite{lauga2009a,elgeti2015}. From the beating pattern of a single cilium \cite{osterman2011,eloy2012} to the body shape of a microorganism \cite{vilfan2012}, efficiency has been shown to be a key factor in better understanding the realm of microorganisms. 

Efficiency is also important for artificial swimmers. While for diffusiophoretic swimmers efficiency is often discussed from thermodynamical perspectives \cite{sabass2010,sabass2012,pietzonka2016}, for swimmers that rely on body deformations (or strokes) for locomotion, swimming (or hydrodynamic) efficiency is used as a measure for optimizing the propulsion mechanisms \cite{avron2004,tam2007,michelin2011}. Specifically, energy dissipation scales quadratically with speed, and so fast propulsion necessitates optimal swimming strokes. These strokes, to be able to have any net effect, must be nonreciproal: they should not remain unchanged under a time reversal. This constraint, colloquially referred to as the scallop theorem \cite{purcell1977}, stems from the absence of inertia and the consequent linearity of the field equations. Within the past few decades, several theoretical models have been proposed that can evade this constraint, either with the aim of explaining the swimming strategies of microorganisms \cite{taylor1951,lighthill1952} or finding the simplest techniques for locomotion (see \cite{lauga2011} for a review). Among these, the Najafi-Golestanian three-sphere swimmer has proven quite useful in manifesting a simple approach for creating a nonreciprocal motion, using only two degrees of freedom \cite{najafi2004}. The swimmer consists of three equally-sized spheres connected by two vanishingly-thin rods. This collinear assembly propels itself forward following a four-step motion wherein the rods change their length between two prescribed values in a nonreciprocal fashion. After a full cycle, the swimmer thereby returns to its original configuration, but has translated forward \cite{golestanian2008}. Owing to its simplicity, many have adapted/extended this model swimmer to further investigate the propulsion mechanisms: from the experimental realization \cite{leoni2009,grosjean2016}, to including near-field effects of a finite sphere size \cite{golestanian2008b}, to introduction of elasticity (in the rods \cite{pande2015,montino2015}, spheres \cite{nasouri2017}, or fluid \cite{datt2018}), an adjacent wall \cite{zargar2009,daddi2018}, reinforcement algorithms \cite{tsang2018}, or even inertia \cite{klotsa2015,felderhof2016,dombrowski2019}. The stochastic adaptation of this swimmer has also been used to describe the molecular swimming \cite{golestanian2008c,golestanian2009,golestanian2010} and the motion of enzymes \cite{golestanian2015,bai2015}.

The efficiency of the three-sphere swimmer with minimal arm-length variations (as originally proposed \cite{najafi2004}) is relatively low \cite{avron2005}, and so optimizing this swimmer has also attracted some attention \cite{felderhof2015}. In a series of studies \cite{alouges2007,alouges2009,alouges2011}, Alouges and coworkers looked into optimal strokes for axisymmetric swimmers at low Reynolds number, taking the three-sphere swimmer as one of their primary models. By casting the optimization problem in the language of control theory, they discussed the optimal strokes for the swimmer with equally-sized spheres, for which the translation per period is prescribed. In this work, we intend to relax this constraint and look at the maximum feasible efficiency of the three-sphere swimmer when it is allowed to select not only its strokes, but also the radii of its spheres. By accounting for the full hydrodynamic interactions, we show that there exists a specific size ratio of the spheres at which the swimmer has the highest efficiency. We also show that the value of this size ratio remains unchanged when we vary the maximum permissible arm length. We finally look at a limiting case to determine the maximum efficiency that the three-sphere swimmer can ever reach, and then use it to prove that the optimal size ratio holds even when the swimmer is allowed to extend its body indefinitely.

The paper is organized as follows: we first introduce the generalized three-sphere swimmer model wherein the spheres can be of different sizes and the arms can change their length from zero to a maximum that we prescribe. Then after solving the full hydrodynamic interactions using a boundary element method, we optimize the swimming strokes and evaluate the variations of the swimming efficiency with respect to the size ratio of the spheres. We then look into the effects of the maximum allowable arm lengths on the swimming efficiency, and finally introduce a limiting case that can accurately set the upper limit for the efficiency of the three-sphere swimmer.  
\section{Governing Equations}
\subsection{Model swimmer}
\begin{figure}[H]
\begin{center}
\includegraphics[width=0.6\textwidth]{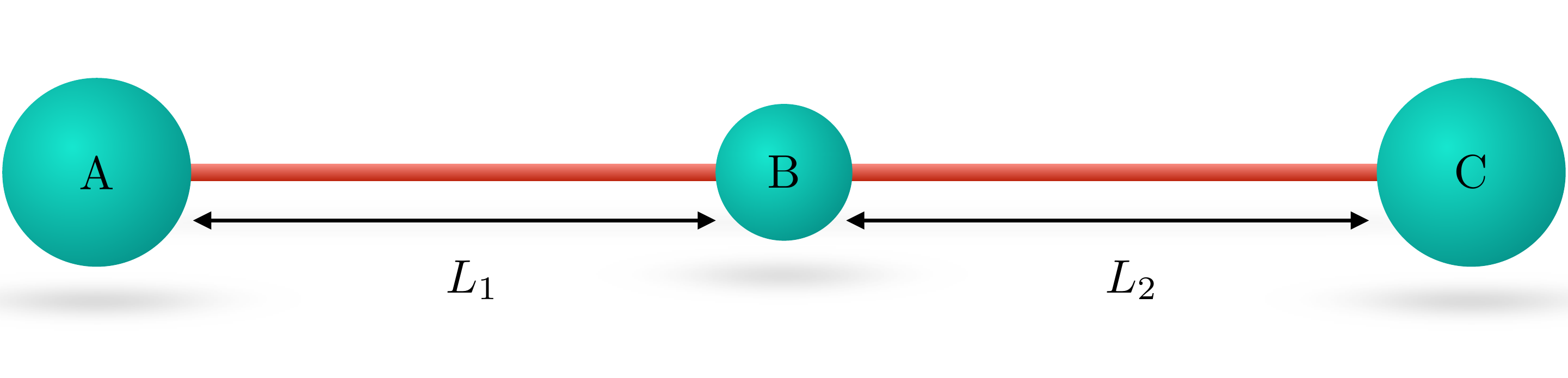}
\caption{Schematic of the three-sphere swimmer considered in this study. The swimmer can propel by opening and closing its arms, and we refer to their instantaneous lengths as $L_1$ and $L_2$.}
\label{model}
\end{center}
\end{figure}

We consider a swimmer comprising of three spheres connected by two rods in an otherwise quiescent viscous fluid of viscosity $\mu$, as shown in Fig.~\ref{model}. The spheres on the sides are of radii $r_1$ and the radius of the middle sphere is $r_2$. 
The rods, which are hydrodynamically unimportant, are of lengths $L_1$ and $L_2$, and can vary their length from zero to $L_{\text{max}}$. The arms undergo periodic length variations (i.e., strokes) which lead to propulsion of the whole body, provided they are nonreciprocal. Here, we want to find the specific strokes and size ratio (i.e., $r_1/r_2$) that optimize the propulsion mechanism of the swimmer. For the concept of optimal swimming, we follow the classical work of \citet{lighthill1952}, in which the efficiency is defined as the ratio of the \textit{required} power to pull the swimmer at a certain speed and the actual expended one during active locomotion with the same average speed.

As the first step, we look into the hydrodynamic interactions between the spheres and then use that to determine the efficiency. At any instant of the motion, the rods apply forces $f_A\boldsymbol{e}$, $f_B\boldsymbol{e}$ and $f_C\boldsymbol{e}$ on the spheres A, B, and C, where $\boldsymbol{e}$ is a unit vector representing the axis of symmetry. Relying on the linearity of the Stokes equations and defining ${U}_A\boldsymbol{e}$, ${U}_B\boldsymbol{e}$, and ${U}_C\boldsymbol{e}$ as the instantaneous velocities of the spheres, we may write the drag law as
\begin{align}
\label{drag_law}
\left[\begin{matrix} 
f_A \\
f_B\\
f_C
\end{matrix}\right]=\left[\begin{matrix} 
R_{AA} &R_{AB} & R_{AC}\\
R_{BA} &R_{BB} & R_{BC}\\
R_{CA} &R_{CB} & R_{CC}
\end{matrix}\right]\,
\left[\begin{matrix} 
U_A \\
U_B\\
U_C
\end{matrix}\right],
\end{align}
or in short $\boldsymbol{F}=\boldsymbol{R}\,\boldsymbol{U}.$
Note that here, $\boldsymbol{R}$ is the resistance tensor mapping translational velocities to forces and is solely a function of the relative positions of the spheres and their sizes. 
We assume that the swimmer can indeed swim without the presence of any external force, and recalling that the motion is overdamped, one can claim 
\begin{align}
\label{force_free}
f_A+f_B+f_C=0.
\end{align}
We can close the equations by defining $\dot{L}_1\boldsymbol{e}$ and $\dot{L}_2\boldsymbol{e}$ as the relative velocities of the spheres, and so we have
\begin{align}
\label{rel_vel}
U_B-U_A=\dot{L}_1,\quad U_C-U_B=\dot{L}_2.
\end{align}
We have now arrived at a system of six equations (Eqs.~\eqref{drag_law} to \eqref{rel_vel}) with six unknowns ($\boldsymbol{F}$ and $\boldsymbol{U}$). Thus, we can explicitly find the instantaneous forces and velocities of the spheres in terms of $\dot{L}_1$ and $\dot{L}_2$, provided the resistance tensor is known. Once solved over a full cycle, we can also find the net translation of the system. Defining $\dot{\boldsymbol{L}}=\left[\dot{L}_1,\dot{L}_2\right]$, the solution takes the form
\begin{align}
\boldsymbol{U}=\boldsymbol{\mathcal{C}}\,\dot{\boldsymbol{L}},
\end{align}
where the $3\times 2$ matrix $\boldsymbol{\mathcal{C}}$ is a function of components of $\boldsymbol{R}$, and can be explicitly found from Eqs.~\eqref{drag_law} to \eqref{rel_vel}.
Now, the Lighthill efficiency reads
\begin{align}
\label{eff}
\eta=\frac{6\pi\mu r_\text{eff} \bar{U}^2}{\mathcal{P}},
\end{align}
where the expended power by the swimmer 
\begin{align}
\label{power}
\mathcal{P}=\frac{1}{T}\int_{0}^{T}\boldsymbol{U}\cdot\boldsymbol{F}\text{d}t=\frac{1}{T}\int_{0}^{T}\left(\dot{\boldsymbol{L}}^\top\,\boldsymbol{\mathcal{C}}^\top\right)\,\boldsymbol{{R}}\,\left(\boldsymbol{\mathcal{C}}\,\dot{\boldsymbol{L}}\right)\text{d}t,
\end{align}
and $T$ is the period. 
The average speed of the swimmer can be defined as the net translation of any reference point on the swimmer (e.g., sphere B) per one complete cycle. Thus, we may define it as 
\begin{align}
\bar{U}=\frac 1 T \int_0^T \left(0,1,0\right)^\top\cdot\boldsymbol{\mathcal{C}}\,\dot{\boldsymbol{L}}\, \text{d}t\,.
\end{align}
Characteristic for the Stokes hydrodynamics, $\bar U$ only depends on the form of the trajectory and the cycle time, but not on the details of the velocity profile through the cycle.

The effective radius $r_\text{eff}$ determines the Stokes drag that is used to calculate the equivalent dissipation, if the swimmer is pulled by an external force with speed $\bar U$. For a swimmer that undergoes large body deformations, such as the three-sphere swimmer considered here, the definition of the swimmer's drag coefficient is not clear. We choose $r_\text{eff}=\sqrt[3]{{2r_1^3+r_2^3}}$ as the radius of a sphere with the same total volume as the three spheres of the swimmer combined. The rationale behind the choice is to use the power needed to transport a spherical cargo of the same volume as a reference. We note that $r_\text{eff}$ only serves as a scaling factor in Eq.~\eqref{eff} and so using different definitions (e.g., $r_\text{eff}=2r_1+r_2$, which represents the combined drag on the three spheres without hydrodynamic interactions) does not alter the results qualitatively. Finally, it is important to note that the Lighthill efficiency is independent of period $T$ and the swimmer size, as one can simply show from \eqref{eff}. Therefore, the length variations profile and the size ratio uniquely determine the efficiency. We also note that unlike the case of cylindrical swimmers, the Lighthill efficiency provides a reliable measure for calculating the efficiency of sphere-based swimmers, such as the three-sphere swimmer considered here \cite{shapere1989}. By maximizing the Lighthill efficiency, we will always find the way to move a swimmer with a given total volume at a given speed with minimal power. Conversely, it will also give us the maximum swimming speed a swimmer with a given volume and available power can achieve.

\subsection{Computational framework}
We discretize the length variations of the rods during one cycle using $N$ time points, and at each segment, refer to their lengths using $\boldsymbol{L}_i=\left[L_1^i,L_2^i\right]$, where $i\in\{1,2,\cdots,N\}$. Our aim is to find a trajectory (i.e., $\boldsymbol{L}_i$) and a size ratio ($r_1/r_2$) that maximizes the efficiency of the swimmer. Under this discretization, we can calculate the translational speed and the net power as
\begin{align}
\label{U2}
\bar{U}&=\frac{1}{2N\Delta t}\sum_{i=1}^N \left[\left(0,1,0\right)^\top\cdot\left(\boldsymbol{\mathcal{C}}_i+\boldsymbol{\mathcal{C}}_{i+1}\right)\right]\,\left({\boldsymbol{L}}_{i+1}-{\boldsymbol{L}}_i\right),\\
\label{P2}
  \mathcal{P}&=\frac{1}{2N\Delta t^2}\sum_{i=1}^N\left({\boldsymbol{L}}_{i+1}-{\boldsymbol{L}}_i\right)^{\top}\left({\boldsymbol{\mathcal{C}}}_i^{\top} \boldsymbol{{R}}_i {\boldsymbol{\mathcal{C}}}_i   +
{\boldsymbol{\mathcal{C}}}_{i+1}^{\top} \boldsymbol{{R}}_{i+1} {\boldsymbol{\mathcal{C}}}_{i+1} \right)
             \left({\boldsymbol{L}}_{i+1}-{\boldsymbol{L}}_i\right),
\end{align}
where the index boundaries are periodic (e.g., $\boldsymbol{L}_{N+1}\equiv \boldsymbol{L}_1$), and $\Delta t=T/N$ is the time step. 

To find $\boldsymbol{R}$ (and subsequently $\boldsymbol{\mathcal{C}}$), we solve for the hydrodynamic interactions between the spheres by adapting a boundary element method routine {\texttt{prtcl\_ax}} from the BEMLIB library by \citet{pozrikidis2002}. We validate our results against those obtained using HYDROLIB \cite{hinsen1995} for the case of equally-sized spheres, and observe excellent agreement. To proceed with the optimization, by means of Numerical Algorithm Group (NAG) routine {\texttt{e04jyf}}, we use a quasi-Newton algorithm for tuning $2N$ variables of $\boldsymbol{L}_i$. For a given set of $\boldsymbol{L}_i$ generated by the optimizer, the efficiency is calculated by substituting \eqref{U2} and \eqref{P2} into \eqref{eff}, and the iterations continue until no higher value can be reached. We repeat the procedure systematically for different values of size ratio and maximum allowable arm length ($L_\text{max}$). For the results reported henceforth, we have set $N=100$, as further increasing $N$ did not result in any quantitative changes. 

\section{Results and Discussion}
\begin{figure}
\begin{center}
\includegraphics[width=\textwidth]{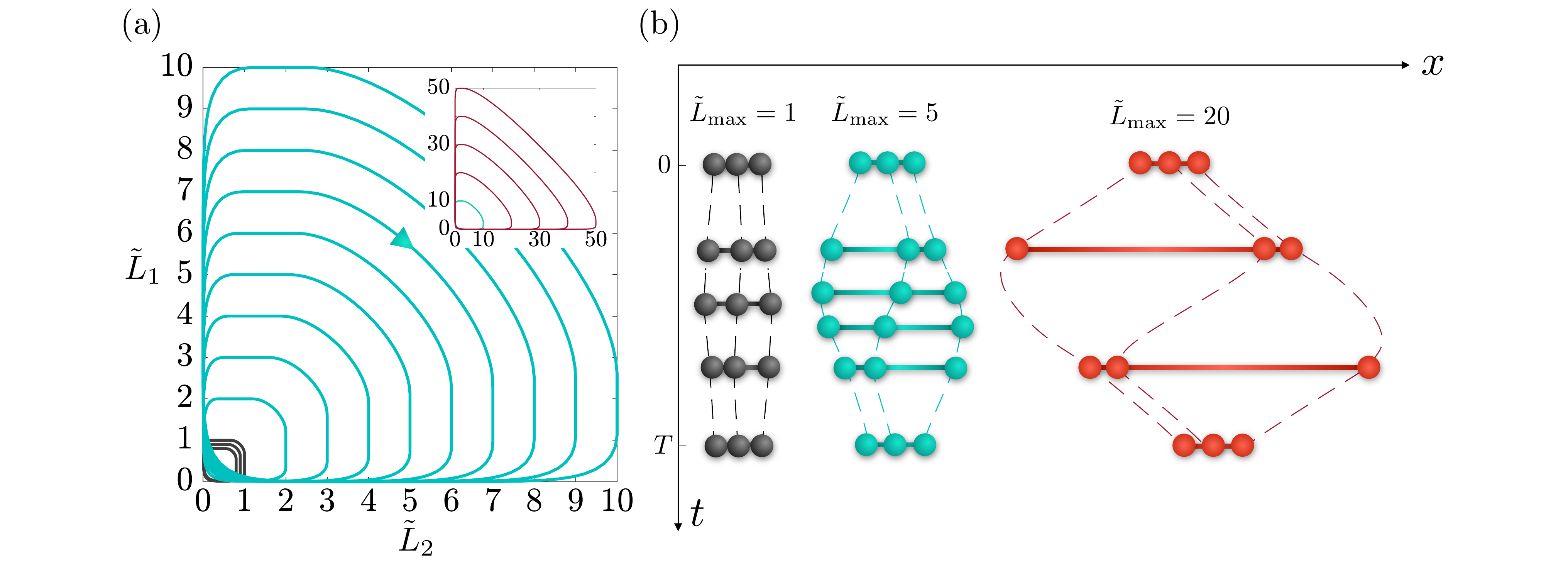}
\caption{(a) Optimal strokes of the three-sphere swimmer, for different values of maximum allowable arm length $\tilde{L}_\text{max}$. $\tilde{L}_1$ and $\tilde{L}_2$ are the arm lengths, and the swimmer follows these trajectories clockwise. Different classes of trajectories are marked by different colors and the trajectories for $20\lesssim\tilde{L}_\text{max}$ are shown in inset. (b) Examples of the optimal propulsion mechanisms for three different classes of the trajectories versus time. The dashed lines indicate the position of each sphere throughout a cycle and colors correspond to trajectory classes.}
\label{trajec}
\end{center}
\end{figure}

To properly evaluate the role of sphere radii on the efficiency, throughout this section, we scale lengths with $r_\text{ch}=\sqrt[3]{(1/3)\left({2r_1^3+r_2^3}\right)}$ and use notation ( $\tilde{}$ ) to distinguish dimensionless entities from dimensional ones. We begin our analysis with the optimal trajectory of the swimmer. We find that the optimal trajectories can be broadly categorized into three classes, depending on the value of $\tilde{L}_\text{max}$. For $\tilde{L}_\text{max}\lesssim1$, the swimmer follows a four-step motion, shown by black squares in Fig.~\ref{trajec}(a). Starting from all three spheres touching, the sequence goes as follows: The left arm opens first, the right arm opens so the swimmer reaches its fully extended state, then the left arm closes and finally the right arm closes, as illustrated in Fig.~\ref{trajec}(b). When $1<\tilde{L}_\text{max}<20$, the fully extended state is no longer sought, and the swimmer rather takes two short steps, resulting in a semi-trapezoidal trajectory shown in Fig.~\ref{trajec} (those in blue color). Interestingly, further increasing the maximum allowable arm length ($20\lesssim\tilde{L}_\text{max}$) causes the swimmer to select a simpler trajectory consisting of three steps: opening of the left arm, then opening of the right arm while simultaneously closing the left arm, and finally closing of the right arm (trajectories in red). 

To elude the scallop theorem, the swimmer exploits the nonreciprocal asymmetries it can create with its cyclic body deformations, and so ``stronger" asymmetries can lead to larger net displacements. Our results indicate that indeed the swimmer maximizes its asymmetry by first extending one arm and closing the other, and then vice versa. In fact, these two states of the swimmer are shared between all classes of the optimal trajectories shown in Fig.~\ref{trajec}, and the only difference between these trajectories arises in the steps the swimmer takes between these two common states. 

\begin{figure}
\begin{center}
\includegraphics[width=\textwidth]{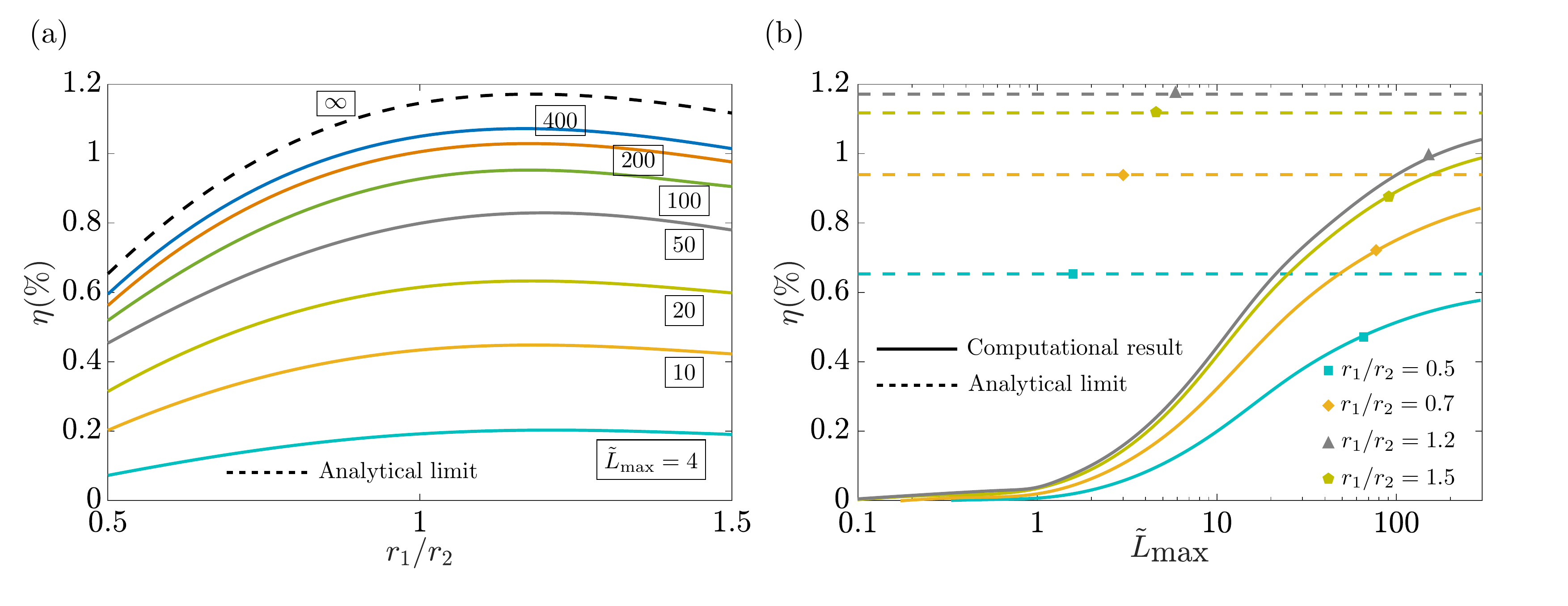}
\caption{(a) Variations of the efficiency with the spheres size ratio, for different values of maximum allowable arm length $\tilde{L}_\text{max}$. The dashed line indicates the efficiency for the limiting case at which $\tilde{L}_\text{max}\rightarrow\infty$. (b) Efficiency versus maximum allowable arm length for different size ratios. Solid lines are the computational results and dashed lines are those of the limiting case. Colors (and symbols) refer to size ratios. See ancillary files for races between swimmers with different size ratios and maximum allowable arm lengths.}
\label{result}
\end{center}
\end{figure}

We may now determine the effect of size ratio of the spheres on the efficiency of the swimmer. Clearly, when $r_1/r_2\rightarrow\infty$, we are making the middle sphere obsolete, essentially removing one degree of freedom from the swimmer. Hence, the efficiency is negligibly small at this limit. Similarly, $r_1/r_2\rightarrow 0$ is inefficient as in this case both degrees of freedom vanish. Thus, the variation of the efficiency with size ratio is nonmonotonic, as our computational results also confirm in Fig.~\ref{result}(a). 
Remarkably, for all values of $L_\text{max}$, the efficiency reaches its maximum when the size ratio is $r_1/r_2=1.196\approx 1.2$. Thus, there exists a unique size ratio at which the swimmer is optimal, and that is when the side spheres are around $20\%$ larger than the middle one.



On the other hand, as expected, efficiency monotonically rises with increasing the maximum allowable arm length, as shown in Fig.~\ref{result}(b). As $L_\text{max}$ increases, the swimmer can further extend its body, allowing for more propulsive thrust. Thus, $\bar{U}$ rises, but so does the viscous dissipations. While the increase of the former is shown to be always faster than the latter, the efficiency seems to plateau for very large values of $\tilde{L}_\text{max}$. Indeed, by allowing for larger body extensions, we cannot increase the efficiency of the swimmer indefinitely, and there exists an upper limit that the efficiency tends toward when $\tilde{L}_\text{max}\rightarrow\infty$. To find this upper limit, we consider a swimmer for which there are no length constraints on the arms. As shown earlier, this swimmer has a three-step motion, depicted in Fig.~\ref{trajec}(b). At the first step, sphere A moves away from spheres B and C. Since $\tilde{L}_\text{max}$ is infinitely large and that hydrodynamic interactions decay quadratically with distance \cite{happel1981,nasouri2018}, one may neglect the hydrodynamic interactions between sphere A and the other two spheres. Under this approximation, and recalling that spheres B and C do not have any relative motion in this step, we can alternatively formulate the drag law as 
\begin{align}
\label{drag_law2}
\left[\begin{matrix} 
f_A \\
f_{(BC)}
\end{matrix}\right]=\left[\begin{matrix} 
R_{AA} &0 \\
0 &R_{(BC)(BC)} 
\end{matrix}\right]\,\left[\begin{matrix} 
U_A \\
U_{(BC)}
\end{matrix}\right].
\end{align}
Here, we have grouped spheres B and C as a single particle, namely particle (BC), and diagonalized the resistance tensor by ignoring hydrodynamic interactions between assembly (BC) and sphere A. In the second step, sphere B separates from sphere C and moves toward sphere A. Relying on the same argument provided for the first step, this time, we neglect the hydrodynamic interactions between all the spheres and arrive at
\begin{align}
\label{drag_law3}
\left[\begin{matrix} 
f_A \\
f_{B}\\
f_C
\end{matrix}\right]=\left[\begin{matrix} 
R_{AA} &0 &0\\
0 &R_{BB}&0\\
0&0&R_{CC} 
\end{matrix}\right]\,\left[\begin{matrix} 
U_A \\
U_{B}\\
U_C
\end{matrix}\right].
\end{align}
Finally, for the third step, we group spheres A and B and so the drag law is reduced to
\begin{align}
\label{drag_law4}
\left[\begin{matrix} 
f_{(AB)} \\
f_{C}
\end{matrix}\right]=\left[\begin{matrix} 
R_{(AB)(AB)} &0 \\
0 &R_{CC} 
\end{matrix}\right]\,\left[\begin{matrix} 
U_{(AB)} \\
U_{C}
\end{matrix}\right].
\end{align}
These equations, combined with Eqs.~\eqref{force_free} and \eqref{rel_vel}, can simply be solved and the net displacement and dissipated power at each step can be found. Note that here the resistance tensors are exactly known. For the ``isolated" spheres, we have $R_{AA}=R_{CC}=6\pi\mu r_1$, and $R_{BB}=6\pi\mu r_2$. For the two-sphere assemblies, we use the exact solutions for motion of two spheres in Stokes flow, obtained by \citet{stimson1926}, and \citet{maude1961} (also see \cite{spielman1970} for some corrections). For instance, when $r_1=r_2$, we have $R_{(AB)(AB)}=R_{(BC)(BC)}\approx7.74\pi\mu r_1$. Now to find the efficiency in this limiting case, we need to determine the net displacement and the net power. However, one cannot simply add the contribution of these three steps, as the period may not be equally divided between them; the swimmer may spend more time in a step compared to another. Relying on the fact that the rate of energy dissipation must be constant throughout an optimal stroke \cite{alouges2009}, we simply find the portion of the period spent on each step, use that to find the net displacement and power, and finally calculate the efficiency. We see that this limiting case indeed sets the upper limit for the efficiencies, as shown by dashed lines in Fig.~\ref{result}. Interestingly, the optimal size ratio of $\sim1.2$ is also accurately captured in this analytical limit.

\section{Conclusion}

We have determined the efficiency limits and optimal strokes for the three-sphere swimmer with no other constraints than the maximal allowed arm extension. This result gives us the minimum power needed for the locomotion of a three-sphere swimmer with a given total volume at a given speed. Whereas the previous studies assumed spheres of equal sizes, we show that the efficiency can be improved if the side spheres are $20\%$ larger than the central one.  Depending on the allowed extension, the optimal trajectories change their shape from a 4-point to a 3-point cycle. We can finally state that to have the most efficient three-sphere swimmer, the arms should be able to extend indefinitely and they should follow a three-step motion shown in Fig.~\ref{trajec}. All these combined, the absolute upper limit for the three-sphere swimmer is found to be $1.17\%$. It is interesting to note that the efficiency limit of the three-sphere swimmer is of the same order of magnitude as that of a ciliated microorganism \cite{osterman2011}, a bacterium with a rotary flagellum \cite{purcell1977} and somewhat lower than the efficiency of beating flagella ($6\,\%$ \cite{spagnolie2010b}).

\section*{Acknowledgement}
 A.V. acknowledges support from the Slovenian Research Agency (grant no. P1-0099).

\bibliography{reference}

\end{document}